\documentclass{moriond}

\def\Journal#1#2#3#4{{#1} {\bf #2}, #3 (#4)}

\def\PRL{\em Phys. Rev. Lett.}
\def\PRD{{\em Phys. Rev.} D}

\def\be{\begin{equation}}
\def\ee{\end{equation}}
\def\bea{\begin{eqnarray}}
\def\eea{\end{eqnarray}}

\usepackage[separate-uncertainty]{siunitx}
\DeclareSIUnit\years{yrs}
\DeclareSIUnit\photoelectron{PE}
\DeclareSIUnit\parsec{pc}

\usepackage{lineno}

\newcommand{\Photo}{}

\begin{document}
\vspace*{4cm}
\title{Recent cosmogenic neutrino search results with IceCube and prospects with IceCube-Gen2}

\author{Maximilian Meier for the IceCube collaboration \footnote{\href{http://icecube.wisc.edu}{icecube.wisc.edu}}}

\address{International Center for Hadron Astrophysics, Chiba University \\
263-8522 Chiba, Japan}

\maketitle\abstracts{
Neutrinos with energies beyond PeV (extremely high energy, EHE) are produced in interaction of the highest energy cosmic rays. One contribution to the EHE neutrino flux is expected to arise from so-called cosmogenic neutrinos generated in ultra high energy cosmic ray interactions with cosmic microwave background photons. Observation of cosmogenic neutrinos can probe the nature of cosmic rays beyond the energies for resonant photo-pion production (GZK cutoff). The IceCube detector instruments a cubic kilometer of the South Pole ice to detect Cherenkov light emitted by charged particles produced in neutrino interactions. In the future IceCube-Gen2 will increase the effective detection volume for EHE neutrinos by adding radio antennas to the in-ice detector. In this contribution we present new constraints on the EHE neutrino flux above \SI{5e6}{\giga\electronvolt} using \num{12.6} years of IceCube data. The differential upper limit constrains the all-flavor EHE neutrino flux at \SI{1}{\exa\electronvolt} to below a level of $E^2 \Phi \sim \SI{e-8}{\giga\electronvolt \per\centi\metre\squared \per\second \per\steradian}$. Additionally, we also describe the projected sensitivity of the IceCube-Gen2 radio array, which will reach fluxes about 1.5 orders of magnitude smaller than the current limits at \SI{1}{\exa\electronvolt}.}

\section{Introduction}
Extremely high energy (EHE) neutrinos are unique messengers of the high redshift universe. They can reach Earth basically unattenuated from great distances ($\mathcal{O}(\SI{50}{\mega\parsec})$). Ultra high energy cosmic rays (UHECR) are deflected by intergalactic magnetic fields and their flux at energies above $\sim\SI{3e19}{\electronvolt}$ is suppressed by the so-called Greisen-Zatsepin-Kuzmin (GZK) mechanism~\cite{greisen}$^{,}$~\cite{zatsepin_kuzmin} through interactions with the cosmic microwave background (CMB). High energy gammas rays are also strongly attenuated by interactions with the extragalactic background light and the CMB. On the other hand, neutrinos are neutral particles and only interact weakly. A flux of high energy neutrinos is expected to be produced by the aforementioned GZK interactions of the proton component of UHECRs during their propagation through the universe, called \textit{cosmogenic} neutrinos. Neutrinos can also be produced in the astrophysical sources directly, and will thus be referred to as \textit{astrophysical} neutrinos. Cosmogenic neutrinos hold unique information about the sources of UHECRs. The normalization and the shape of the neutrino flux carries information about the cosmic ray composition, the redshift evolution of their sources and the maximum energy of their accelerators.

The IceCube Neutrino Observatory has previously published searches for this component of the neutrino flux~\cite{ehe_prd}. In this proceeding, we report a new revision of this search including a total of \SI{12.6}{\years} of detector livetime as well as an updated event selection. 

\section{IceCube Detector}
\begin{figure}
    \centering
    \includegraphics[width=.75\textwidth]{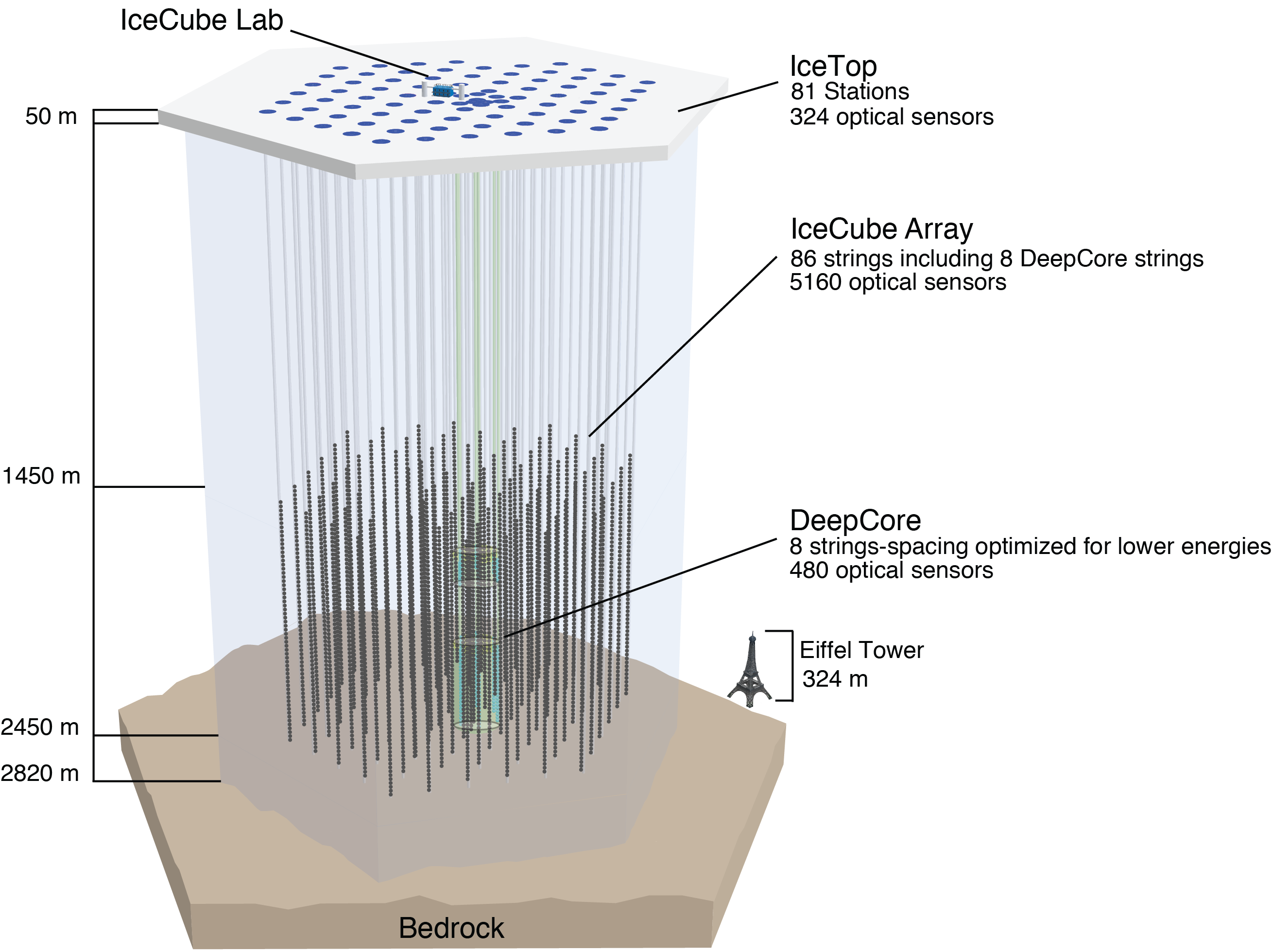}
    \caption{Sketch of the IceCube Neutrino Observatory showing the main in-ice array as well as the surface detector IceTop.}
    \label{fig:ic}
\end{figure}

The IceCube detector instruments a cubic kilometer of glacial ice at the geographic South Pole~\cite{ic_detector}. The in-ice array consists of \num{5160} Digital Optical Modules (DOMs) distributed along 86 strings buried into the ice at depths between \SI{1500}{\metre} and \SI{2500}{\metre} with an interstring spacing of \SI{125}{\metre} (see Fig.~\ref{fig:ic}). Each string hosts \num{60} DOM spaced \SI{15}{\metre} apart vertically. The DOMs detect optical Cherenkov photons with a downward-facing 10 inch PMT. The construction of IceCube was phased, starting in 2007. Until 2010 79 strings were deployed (IC79) and construction was finished in 2011 with the last of the 86 strings getting deployed (IC86).
In addition to the in-ice component, the detector includes the IceTop surface detector to measure extensive air showers induced by cosmic ray interactions with the atmosphere. IceTop consists of two water tanks on top of each string instrumented with two DOMs each.

IceCube observes neutrino-nucleon interactions by detecting Cherenkov photons emitted by secondary charged particles. The PMTs convert these photons to photoelectrons (PE), which are then observed as charge $Q$. IceCube can observe neutrinos from all flavors ($\nu_{\mathrm{e}}$, $\nu_\mu$, $\nu_\tau$) through two main detection topologies, \textit{tracks} and \textit{cascades}. Charged current $\nu_\mu$ and $\nu_\tau$ interactions produce long-range leptons losing their energy stochastically in the detector, appearing as tracks. Charged current $\nu_{\mathrm{e}}$ and all-flavor neutral current interactions produce electromagnetic and/or hadronic showers, which are observed as approximately spherical energy depositions in the detector, resulting in a cascade-like topology. Neutrino interactions producing track-like events can even be detected if they happen far outside of the instrumented volume, enlarging the detectors effective volume by up to about one order of magnitude at the highest energies. In addition to the already mentioned cosmogenic and astrophysical neutrinos, IceCube also detects neutrinos and muons produced in cosmic ray interactions in the atmosphere, which are considered as background events for this search. 

In this analysis, IceCube data from June 2010 to June 2023 is analyzed, which corresponds to an effective livetime of \num{12.6} years of IC86 data. This revision of the analysis is using \textit{Pass2} in-situ detector recalibration~\cite{pass2}, which improves the description of the single-photoelectron charge distribution for the individual PMTs. Compared to the previous search, \num{5.5} years of entirely new data are added, but to treat all of the data consistently with the up-to-date detector calibration, we decided to remove data taken before June 2010 in the partial detector configurations IC22, IC40 and IC59.

\section{Event Selection}
In a search for EHE neutrinos, the main background component in IceCube is atmospheric muons and muon bundles, triggering the detector at a rate of about \SI{3}{\kilo\hertz}. Additional backgrounds are atmospheric neutrinos as well as astrophysical neutrinos. Apart from atmospheric muons, atmospheric neutrinos are the dominant background at low energies, due to their soft energy spectrum with a spectral index of $\gamma \sim -3.7$. Astrophysical neutrinos become the most relevant background component at energies above $\mathcal{O}(\SI{100}{\tera\electronvolt})$, since they have a harder energy spectrum than the other components ($\gamma \sim -2.5$).

The event selection strategy is similar to a previous IceCube study~\cite{ehe_prd}, where signal candidates are identified by successively applying cuts reducing these backgrounds. The different stages are described below.

\subsubsection*{Level 2}
The first stage of the event selection rejects event with a total charge of $Q_{\mathrm{tot}} < \SI[group-separator = {,}]{27500}{\photoelectron}$ or if the number of hit DOMs $n_{\mathrm{DOMs}} < 100$. The majority of the atmospheric muon and neutrino background are already rejected by this requirement.

\subsubsection*{Level 3: Track quality}
\begin{figure}
    \centering
    \begin{minipage}{0.495\textwidth}
        \includegraphics[width=\textwidth]{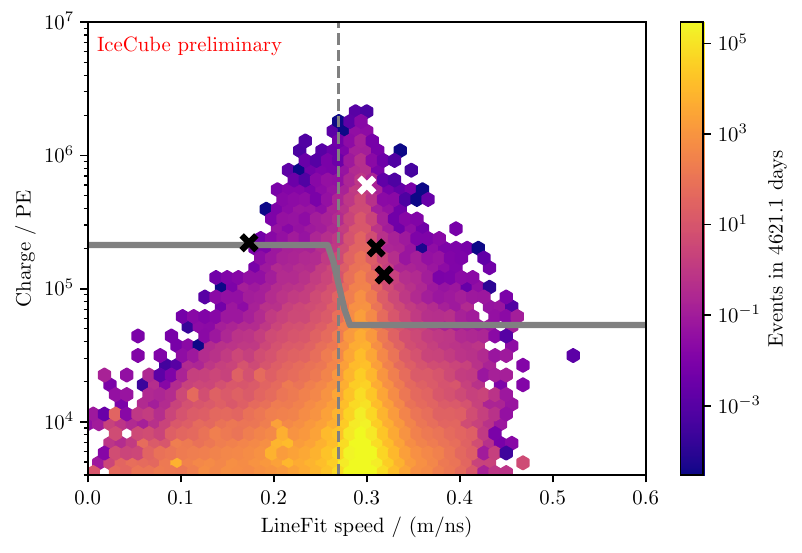}
    \end{minipage}
    \begin{minipage}{0.495\textwidth}
        \includegraphics[width=\textwidth]{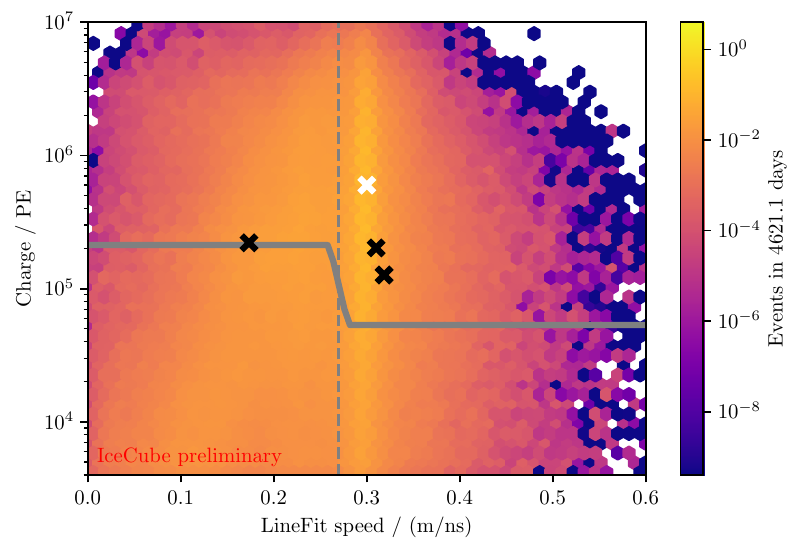}
    \end{minipage}
    \caption{Simulated distribution if LineFit speed and charge for atmospheric muons (left) and cosmogenic neutrinos (right). The events below the grey lines are removed by the Level 3 criterion. The black and white crosses represent the data events passing the event selection up to the Level 4 cut. The event represented by the white cross has been vetoed by the described IceTop veto.}
    \label{fig:l2_cut}
\end{figure}

The track quality cut utilizes the velocity of the \textit{LineFit} reconstruction~\cite{linefit}. The algorithm assumes light propagating as a plane wave with speed $\vec{v}$ along an infinite track. Tracks that are well reconstructed are distributed close to the speed of light $c$, while mis-reconstructed tracks or cascade events are mainly reconstructed to smaller velocities. Because of this property, the LineFit speed can also be used to distinguish between track and cascade topologies (at $v = \SI{0.27}{\metre \per\nano\second}$). The track quality cut is applied as a function of reconstructed speed and charge. The cut, as well as the simulated distributions for atmospheric muons and cosmogenic neutrinos are shown in Fig.~\ref{fig:l2_cut}. The purpose of this cut is the rejection of atmospheric neutrinos and mis-reconstructed events (for example corner-clipping atmospheric muons).

\subsubsection*{Level 4: Muon bundle}
The main purpose of the Level 4 criterion is the removal of the vast majority of down-going muon bundles. We exploit two differences between muon bundles and high energy neutrinos: their energy loss profile and the zenith distribution. The energy loss profile of single muons and taus, which are produced in neutrino interactions differs from the profile expected from muon bundles with large multiplicities. The energy losses of muons/taus become more stochastic with increasing energy. In a muon bundle with the same total energy, its energy is distributed to many muons resulting in a superposition of lower energy muons with more continuous energy losses.
The \textit{stochasticity} of an event can be accessed by a segmented energy loss reconstruction~\cite{millipede}. Then the reconstructed loss profile is compared to an energy loss PDF for muon bundles derived from \texttt{PROPOSAL} simulations~\cite{proposal}. Then a stochasticity proxy of $\sum_i \log(P(\Delta E_i/E) / \mathrm{ndf}$ can be defined similar to a reduced log-likelihood.

\begin{figure}
    \centering
    \begin{minipage}{0.495\textwidth}
        \includegraphics[width=\textwidth]{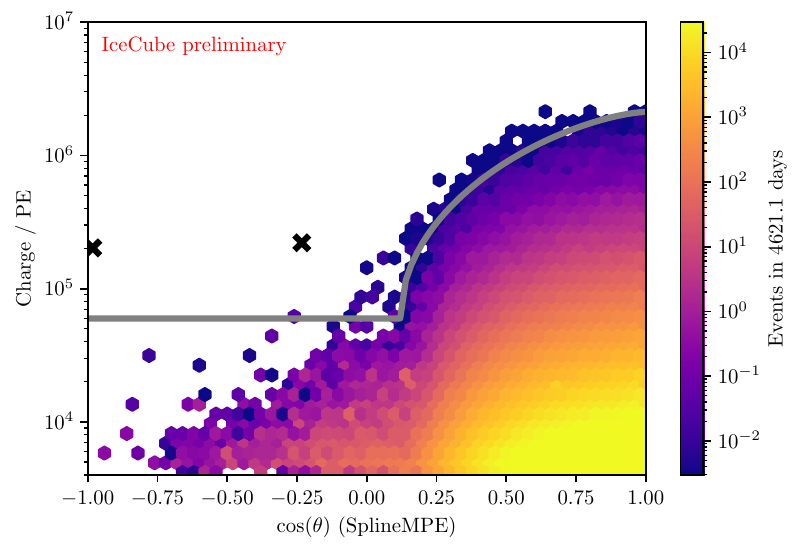}
    \end{minipage}
    \begin{minipage}{0.495\textwidth}
        \includegraphics[width=\textwidth]{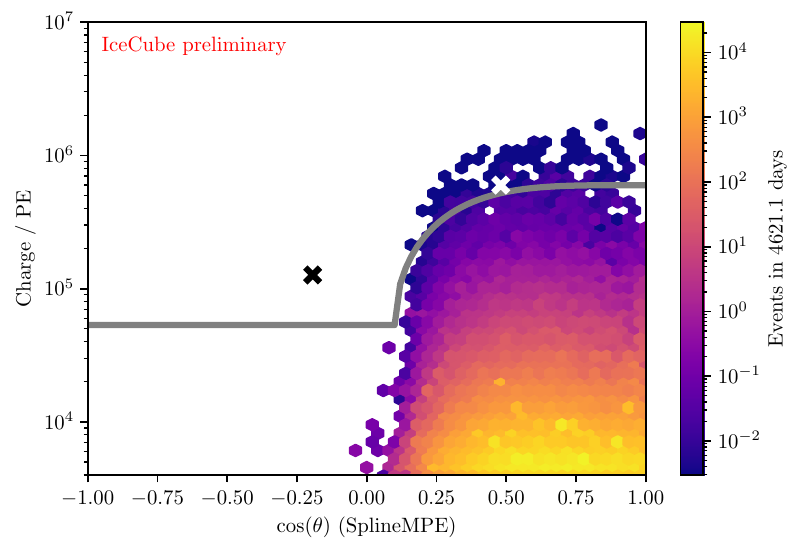}
    \end{minipage}
    \caption{Simulated atmospheric muon background as a function of charge and reconstructed $\cos(\theta)$ for small stochasticities (left) and large stochasticities (right). The events below the grey lines are removed by the Level 4 criterion. The black and white crosses represent the data events passing the event selection up to the Level 4 cut. The event represented by the white cross has been vetoed by the described IceTop veto.}
    \label{fig:l4_cut}
\end{figure}

The muon bundle cut is defined in the 2D plane of reconstructed zenith and charge as shown in Fig.~\ref{fig:l4_cut}. The stochasticity proxy is used to split the sample into sub-samples with low stochasticity and high stochasticity. This allows to loosen the cut in the down-going region ($\cos(\theta) > 0$), which increases the signal efficiency relative to a version of the analysis without the utilization of the stochasticity information.

\subsubsection*{Level 5: IceTop veto}
The surface detector IceTop can be used to further reduce the rate of downgoing atmospheric muons. The in-ice track reconstruction can be used to identify hits in IceTop that are correlated with an in-ice event. A hit in IceTop is considered as a correlated hit if the extrapolated time of closest approach $t_{\mathrm{CA}}$ satisfies the condition: $\SI{-1}{\micro\second} \leq t_{\mathrm{CA}} \leq \SI{1.5}{\micro\second}$. The atmospheric muon background can be reduced by an additional \SI{60}{\percent} by vetoing in-ice events that have two or more correlated IceTop hits. The impact on the all-sky neutrino rate is less than \SI{5}{\percent}.

\hspace{1cm}

\begin{figure}
    \centering
    \includegraphics[width=0.75\textwidth]{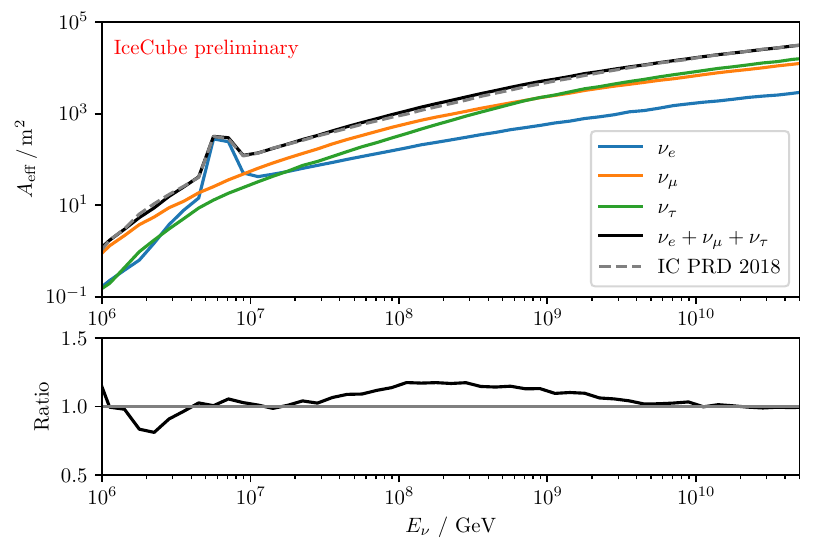}
    \caption[]{Effective area (zenith-averaged) for each neutrino flavor and for the all-flavor sum. Additionally, the all-flavor sum is compared to the all-flavor sum of the previous event selection used by IceCube's cosmogenic neutrino search~\cite{ehe_prd}.}
    \label{fig:aeff}
\end{figure}

The zenith-averaged effective area resulting from the event selection is shown in Fig.~\ref{fig:aeff}. At the highest energies the neutrino sample is clearly dominated by track-like events from $\nu_\mu$ and $\nu_\tau$. The improvements that are made to the event selection improve the $\nu_\mu$ effective area between \SI{100}{\peta\electronvolt} and \SI{1}{\exa\electronvolt} by about \SI{40}{\percent} resulting in a \SI{20}{\percent} improvement of the all-flavor sum. The improvements arise from improved angular reconstruction and the addition of the observed energy loss profiles to the muon bundle rejection. For a detector livetime of \num{12.6} years about \num{0.4} atmospheric background events (including muon bundles and neutrinos) and up to about \num{5} cosmogenic neutrinos for the most optimistic model~\cite{ahlers_gzk} are expected. For the astrophysical background expectation a large range is observed, which arises from the limited knowledge on the behaviour of the astrophysical neutrino flux beyond the \si{\peta\electronvolt}-regime. For an unbroken power law with a hard spectral index~\cite{diffuse_9yr} ($\gamma = \num{-2.37}$) about \num{9} events are expected, while for a power law with a cutoff in the \si{\peta\electronvolt} range~\cite{globalfit} ($\gamma = \num{-2.39}, \,\, E_{\mathrm{cutoff}} = \SI{1.4}{\peta\electronvolt}$) the expectation is reduced to about \num{0.5} astrophysical background events.

After applying the criteria described here to the aforementioned \num{12.6} years of IceCube data, three events are observed. Their reconstructed direction and charge is depicted in Fig.~\ref{fig:l4_cut}. The deposited energy proxy for all events are between \SI{1}{\peta\electronvolt} and \SI{10}{\peta\electronvolt}.

\section{Differential Limit and GZK Model Tests}
The neutrino sample is divided into sub-samples of tracks and cascades as described in the previous section. Then the deposited energy and the arrival direction are reconstructed using likelihood-based reconstructions with a track and a cascade hypothesis respectively. Following the analysis scheme in~\cite{ehe_prd}, the data is fit using a binned Poisson likelihood:
\begin{equation}
    \mathcal{L}(\lambda_{\mathrm{GZK}}, \lambda_{\mathrm{astro}}) = \prod_i P(n_i | \lambda_{\mathrm{GZK}} \mu_{\mathrm{GZK}, i} + \lambda_{\mathrm{astro}} \mu_{\mathrm{astro}, i} + \mu_{\mathrm{bkg}, i}),
    \label{eq:llh}
\end{equation}
where the parameter of interest is the normalization of a cosmogenic neutrino flux model $\lambda_{\mathrm{GZK}}$ and $\lambda_{\mathrm{astro}}$ is a nuisance parameter to describe the normalization of the astrophysical neutrino flux. This analysis is statistically limited and thus can not fulfil the asymptotic conditions for Wilks' theorem. As a result, all hypothesis tests are performed based on ensembles of pseudoexperiments. The impact of systematic uncertainties is estimated by including random variation of the systematic parameters in each pseudoexperiment. This results in a broadening of the TS distribution, shifting the critical value for a given confidence level. Included parameters are the uncertanities on fluxes of background components, the detection efficiency of the DOMs, and the neutrino cross section and inelasticity.

The compatibility of the observed data with a cosmogenic flux model is tested with a likelihood ratio test
\begin{equation}
    \Lambda = \log \left( \frac{\mathcal{L}(\hat{\lambda}_{\mathrm{GZK}}, \hat{\lambda}_{\mathrm{astro}})}{\mathcal{L}(\lambda_{\mathrm{GZK}} = 1, \hat{\lambda}_{\mathrm{astro}})} \right).
    \label{eq:modeltest}
\end{equation}

In the model tests, the astrophysical model from~\cite{diffuse_9yr} is used and different cosmogenic flux models are tested. The resulting p-values for the likelihood ratio test and the \SI{90}{\percent} CL upper limits are presented in Tab.~\ref{tab:modeltest}. The best fit for $\lambda_{\mathrm{GZK}}$ for all tested models is zero. All models assuming an ultra high energy cosmic ray proton fraction of \SI{100}{\percent} are rejected at \SI{90}{\percent} CL.

To obtain a more model-independent constraint on the EHE neutrino flux a differential upper limit can also be constructed. The same likelihood description described in Eq.~\ref{eq:llh} is used, but for a range of energies $E_c$ an $E^{-1}$ flux with a width of one decade in energy is tested. Then, for each energy $E_c$, \SI{90}{\percent} CL Feldman-Cousins confidence intervals are constructed~\cite{feldmancousins}. For the differential upper limit the most conservative result is obtained using an astrophysical background model with a cutoff~\cite{globalfit}.

\begin{figure}
    \centering
    \includegraphics[width=\textwidth]{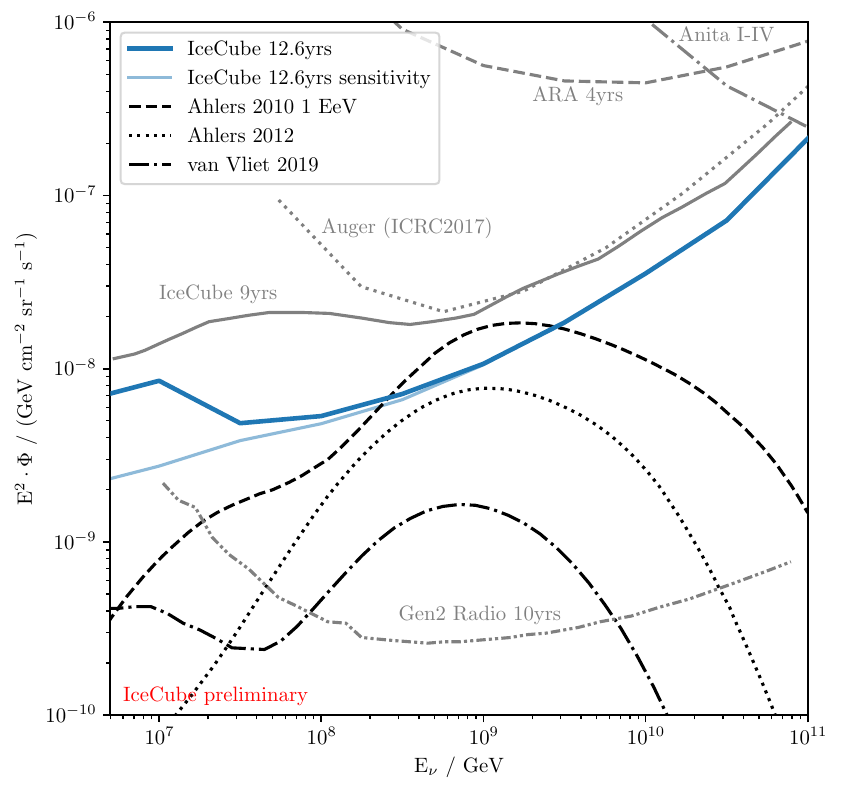}
    \caption[]{Differential upper limit (\SI{90}{\percent} CL)
    on the neutrino  flux for neutrino energies ranging from
    \SI{3}{\peta\electronvolt} to \SI{100}{\exa\electronvolt}.
    The differential limit is compared to a previous IceCube result~\cite{ehe_prd}, limits set by Auger~\cite{auger_limit}, ARA~\cite{ara_limit}, and Anita~\cite{anita_limit} shown as various grey lines, and to cosmogenic neutrino flux models~\cite{ahlers_gzk}$^{,}$~\cite{ahlers_minimal}$^{,}$~\cite{vanvliet}. The model shown from van Vliet et al.~\cite{vanvliet} assumes $\gamma = 2.5$, $E_{\mathrm{max}} = \SI{e20}{\electronvolt}$, $m = 3.4$ and a \SI{10}{\percent} proton fraction.}
    \label{fig:diff_limit}
\end{figure}

The differential limit is shown in Fig.~\ref{fig:diff_limit}. The observed limit (solid blue) is compared to the sensitivity (i.e. the expected limit for a null observation), IceCube's previous limit and limits from other high energy neutrino searches from Auger~\cite{auger_limit}, ARA~\cite{ara_limit} and Anita~\cite{anita_limit}. The black lines show cosmogenic neutrino flux models. The dashed black line from Ahlers et al.~\cite{ahlers_gzk} is the most optimistic model previously not rejected by neutrino observations. The dash-dotted line shows the model from van Vliet et al.~\cite{vanvliet} assuming a proton fraction of \SI{10}{\percent}. The observed limit constraints the all-flavor EHE neutrino flux at \SI{1}{\exa\electronvolt} to a level of $E^2 \Phi < \SI{e-8}{\giga\electronvolt \per\squared\centi\metre \per\second \per\steradian}$.

\begin{table}[]
    \centering
    \caption{Results for the cosmogenic neutrino flux model tests (cf. Eq.~\ref{eq:modeltest}). The first column shows the tested models. The model rejection factor (MRF) in the second column is the upper limit on the flux model relative to its baseline normalization at \SI{90}{\percent} confidence level. The last column is the p-value to reject the model assuming the baseline normalization. The model from van Vliet et al. was tested with the following parameters: $\gamma = 2.5$, $E_{\mathrm{max}} = \SI{e20}{\electronvolt}$, $m = 3.4$ and a \SI{10}{\percent} proton fraction.}
    \begin{tabular}{l||c|c}
        Model & MRF (\SI{90}{\percent}CL UL) & p-value \\
        \hline
        Ahlers et al. 2010 (\SI{1}{\exa\electronvolt})~\cite{ahlers_minimal} & 0.30 & 0.002 \\
        Ahlers et al. 2012~\cite{ahlers_gzk} & 0.58 & 0.032 \\
        Kotera et al. (SFR)~\cite{kotera_sfr} & 0.52 & 0.020 \\
        van Vliet et al.~\cite{vanvliet} & 2.51 & 0.284 \\
    \end{tabular}
    \label{tab:modeltest}
\end{table}

\section{Sensitivity of IceCube Gen2}

A next generation extension of the IceCube neutrino observatory, called IceCube Gen2 is planned~\cite{gen2_tdr}$^{,}$~\cite{gen2_wp}. Scaling up the sensitivity of IceCube using the detection method of optical Cherenkov light is challenging and expensive mainly due to the short attenuation length of $\mathcal{O}(\SI{100}{\metre})$ for optical photons in the ice. To improve the sensitivity to neutrinos with energies above $\sim$\SI{30}{\peta\electronvolt} an additional in-ice radio component is being developed. The radio emission produced by showers induced by high energy neutrino interactions has a larger attenuation length of about \SI{1}{\kilo\metre}. Therefore it allows a sparsely instrumented detector with about \num{360} radio stations to reach an effective volume of $\mathcal{O}(\SI{3600}{\cubic\kilo\metre \steradian})$ at a neutrino energy of \SI{1}{\exa\electronvolt}.

In case of a \SI{10}{\percent} proton fraction of UHECRs IceCube Gen2 will detect more than 3 neutrinos above \SI{100}{\peta\electronvolt} each year. The projected sensitivity to the ultra high energy neutrino flux assuming a livetime of \num{10} years is shown in Fig.~\ref{fig:diff_limit}, reaching an  energy flux level of \SI{3e-10}{\giga\electronvolt \per\squared\centi\metre \per\steradian \per\second}.

\newpage

\section{Conclusion}
In this work, an EHE neutrino search using \num{12.6} years of IceCube data was performed. Three events with reconstructed energies below \SI{10}{\peta\electronvolt} were found, consistent with a non-observation of neutrinos beyond the astrophysical flux measured by IceCube. The differential limit puts the most stringent constraint to date on the cosmogenic neutrino flux above \SI{5e6}{\giga\electronvolt}, reaching an all-flavor neutrino flux of $E^2\Phi \simeq \SI{e-8}{\giga\electronvolt \per\squared\centi\metre \per\steradian \per\second}$ at \SI{1}{\exa\electronvolt}. The performed model tests shows that all tested models assuming a pure proton composition of UHECRs are rejected at \SI{90}{\percent} CL.

In case of a UHECR proton fraction of about \SI{10}{\percent} or less a larger scale neutrino detector is required for a first observation of cosmogenic neutrinos. IceCube-Gen2 will be able to provide experimental constraints for UHECR sources with a mixed composition with a projected sensitivity reaching fluxes about 1.5 orders of magnitude smaller than the current IceCube limit at \SI{1}{\exa\electronvolt}.

\end{document}